\documentclass[lettersize,journal]{IEEEtran}
\usepackage{amsmath,amssymb,amsfonts}
\usepackage{algorithmic}
\usepackage{algorithm}
\usepackage{array}
\usepackage[caption=false,font=normalsize,labelfont=sf,textfont=sf]{subfig}
\usepackage{textcomp}
\usepackage{stfloats}
\usepackage{url}
\usepackage{verbatim}
\usepackage{graphicx}
\usepackage{cite}
\usepackage{tikz}
\usepackage{pgfplots}
\pgfplotsset{compat=1.18} 
\usepackage{xcolor}
\usepackage{microtype}
\usepackage{mathtools}
\usepackage{bbm}
\usepackage{caption}
\numberwithin{equation}{section}

\providecommand{\myparab}[1]{\smallskip\noindent\textbf{#1} }

\def\BibTeX{{\rm B\kern-.05em{\sc i\kern-.025em b}\kern-.08em
    T\kern-.1667em\lower.7ex\hbox{E}\kern-.125emX}}
\usepackage{braket}
\usepackage{hyperref}
\usepackage{physics}
\usepackage{algorithm}
\usepackage{algorithmic}
\usepackage{graphicx}

\providecommand{\myparab}[1]{\smallskip\noindent\textbf{#1} }

\begin{document}
\bibliographystyle{IEEEtran}

\title{Resource Allocation for Entanglement Generation in a Quantum Data Center}
\title{Resource Allocation for Entanglement Generation in a Quantum Data Center}
\title{Entanglement Purification for Fault-Tolerant Distributed Quantum Computing}
\title{A Quantum Data Center for \\ Distributed Quantum Computing}

\title{A Reconfigurable Quantum Data Center Network for \\Distributed Quantum Computing}

\title{Circuit Compilation Techniques for \\Distributed Quantum Computing}
\title{Optimized Quantum Circuit Partitioning and Scheduling for Distributed Quantum Computing}

\title{Scheduler techniques for probabilistic entanglement generation in quantum data centers}
\title{Network-Aware Scheduling for Remote Gate Execution in Quantum Data Centers}




\author{
\IEEEauthorblockN{Shahrooz Pouryousef\IEEEauthorrefmark{1}\IEEEauthorrefmark{2}, Reza Nejabati\IEEEauthorrefmark{2}, Don Towsley\IEEEauthorrefmark{1}, Ramana Kompella\IEEEauthorrefmark{2}, and Eneet Kaur\IEEEauthorrefmark{2}}\\
\IEEEauthorblockA{\IEEEauthorrefmark{1}University of Massachusetts Amherst.
\IEEEauthorrefmark{2}Cisco Quantum Labs, Santa Monica, CA 90404, USA \\
shpouryo@cisco.com
}

}



\maketitle


\begin{abstract}

Modular quantum computing provides a scalable approach to overcome the limitations of monolithic quantum architectures by interconnecting multiple Quantum Processing Units (QPUs) through a quantum network. In this work, we explore and evaluate two entanglement scheduling strategies—static and dynamic—and analyze their performance in terms of circuit execution delay and network resource utilization under realistic assumptions and practical limitations such as probabilistic entanglement generation, limited communication qubits, photonic switch reconfiguration delays, and topology-induced contention. We show that dynamic scheduling consistently outperforms static scheduling in scenarios with high entanglement parallelism, especially when network resources are scarce. Furthermore, we investigate the impact of communication qubit coherence time, modeled as a cutoff for holding EPR pairs, and demonstrate that aggressive lookahead strategies can degrade performance when coherence times are short, due to premature entanglement discarding and wasted resources. We also identify congestion-free BSM provisioning by profiling peak BSM usage per switch. Our results provide actionable insights for scheduler design and resource provisioning in realistic quantum data centers, bringing system-level considerations closer to practical quantum computing deployment.

\end{abstract}

\section{Introduction}

Quantum computing is expected to require thousands of high-fidelity qubits to surpass classical systems on practical tasks \cite{roetteler2017quantumresourceestimatescomputing}. While significant progress has been made in developing various quantum hardware platforms, the challenge of scaling quantum processors remains a critical bottleneck \cite{manetsch2024tweezerarray6100highly, Stephenson2020}. A promising approach to overcoming these limitations is distributed quantum computing (DQC), where quantum circuits are partitioned and executed across multiple quantum processing units (QPUs) \cite{ibm_quantum_roadmap,ionq_quantum_networks,DQC_cisco}.

Once a quantum circuit is partitioned across multiple QPUs, executing non-local gates requires entanglement between distant qubits. This necessitates a network orchestration layer that manages the generation and distribution of EPR pairs across the quantum network. While prior work has proposed techniques such as gate packing \cite{wu2022autocomm,wu2023entanglement}, gate teleportation \cite{Gottesman1999,eisert2000optimal}, and state teleportation \cite{andres2019automated,houshmand2020evolutionary,kaur2025optimized} to reduce the number of non-local gates, the focus of this paper is on the scheduling and orchestration of entanglement generation: given a set of non-local gate dependencies and constrained network resources, how should the orchestrator manage entanglement generation in time and space to minimize circuit execution delay and avoid congestion?



Unlike monolithic quantum systems, DQC introduces unique scheduling constraints arising from the interplay between circuit dependencies (the order of quantum operations) and entanglement dependencies (the timing and success of Bell pair generation). Entanglement generation failures or contention in the network can delay gate execution \cite{DQC_cisco}, forcing the scheduler to make decisions under uncertainty. Moreover, communication qubits have limited coherence times, meaning EPR pairs must be consumed within a certain time window before they decohere—adding another layer of timing pressure to the scheduler's decision-making process. 

\begin{figure}
\centering
    \includegraphics[width=8.8cm]{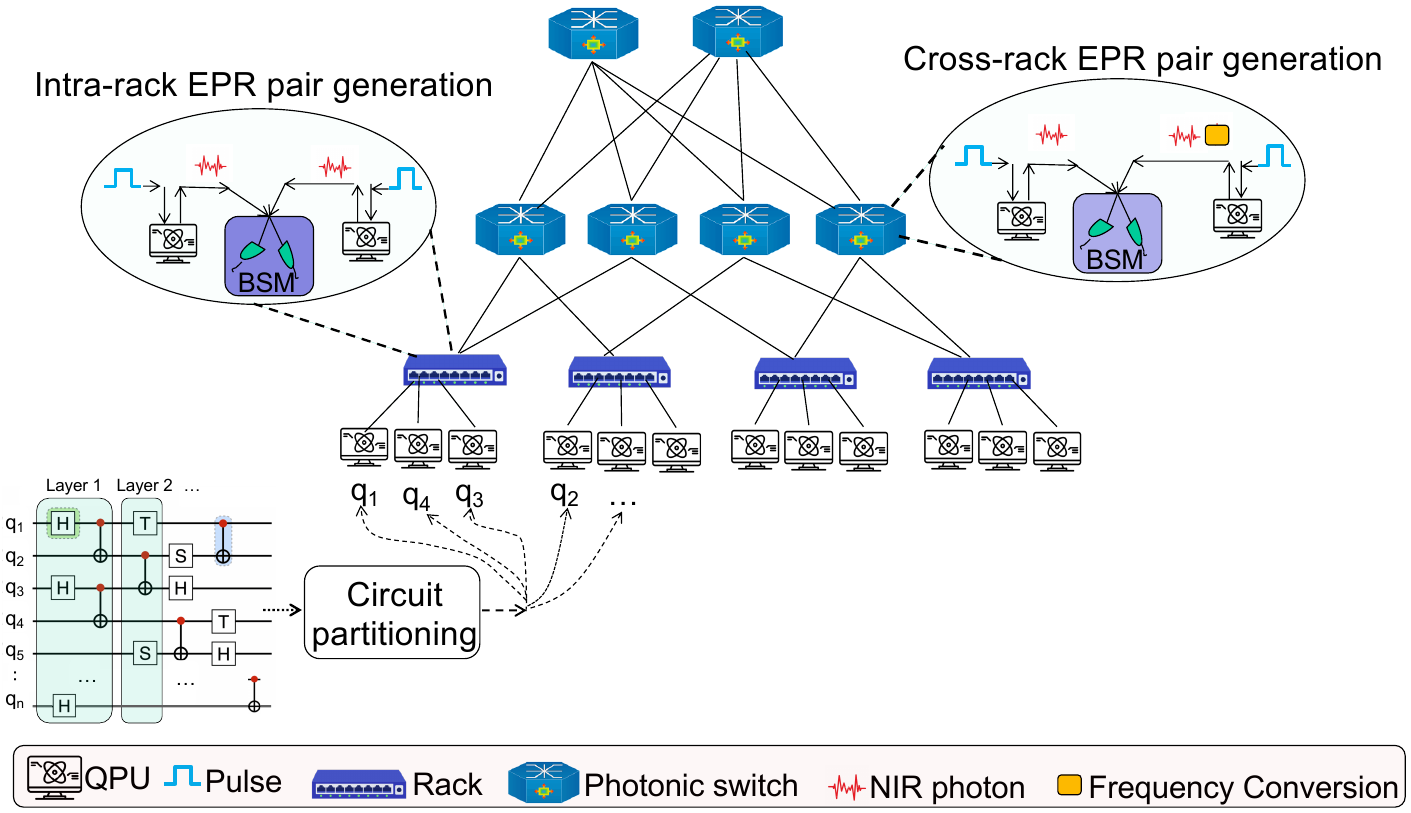}
    \vspace{-0.06in}
\captionsetup{} 
  \caption{A Clos topology for a quantum data center and the physical model of cross-rack and intra-rack entanglement generation.}\label{fig:Clos_cross_rack_intra_rack_pair_gen}
\end{figure}

In this work, we evaluate non-local gate scheduling strategies for executing quantum circuits in a quantum data center environment. Specifically, we compare static and dynamic scheduling approaches under a probabilistic entanglement generation model, taking into account network connectivity and resource constraints. Our primary metric is the total time required to generate all entanglement necessary for the execution of non-local gates. The entanglement process is constrained by limited network resources and the finite coherence time (cutoff) of communication qubits, which limits how long an EPR pair can be stored before it decoheres and must be discarded. In our setup, QPUs are interconnected via reconfigurable photonic switches, enabling dynamic adjustment of connectivity over time. To our knowledge, this is among the first evaluations of non-local gate scheduling that incorporates realistic assumptions about entanglement success probabilities and scalable topologies.

We evaluate performance across a range of benchmark circuits—including QFT, Quantum Volume, QAOA, and Adder—by measuring execution delay under varying entanglement success probabilities within a Clos-topology network (Figure~\ref{fig:Clos_cross_rack_intra_rack_pair_gen}). Our analysis considers both cross-rack and intra-rack entanglement generation protocols between QPUs, reflecting realistic physical layer constraints. Beyond delay, we study the effects of communication qubit coherence time (cutoff limits), analyze QPU-to-QPU entanglement demand heatmaps, and track BSM module utilization over time. Together, these evaluations provide a detailed, architecture-aware understanding of scheduling strategies in quantum data centers and their impact on distributed quantum execution performance. 


Our results show that dynamic scheduling offers significant performance improvements over static scheduling in scenarios with high gate parallelism and sufficient network resources. However, its advantage diminishes for circuits with sparse non-local gates at each layer of the circuit (e.g, QFT circuit) where each layer is the set of independent gates in the circuit. Additionally, we evaluate how cutoff time for storing EPR pairs impacts performance under lookahead scheduling. Our results reveal that overly short coherence windows can lead to premature entanglement discard and degraded performance, while longer cutoffs enable better utilization of early entanglement generation. We also analyze the minimum number of BSMs needed per switch to support congestion-free operation by profiling BSM demand under an idealized unlimited-resource model. These insights provide concrete guidelines for scheduler design and resource provisioning in realistic quantum data centers.

The remainder of this paper is organized as follows. In Section~\ref{sec:schedulers}, we describe the scheduling problem in quantum data centers and introduce the static and dynamic scheduling strategies under realistic network constraints. Section~\ref{sec:eval} presents our evaluation methodology and results across various benchmark circuits and system parameters. Finally, Section~\ref{sec:conclusion} concludes the paper with a summary of key findings.


\section{A Quantum Data Center}
\label{sec:schedulers}

This section describes the architecture of a QDC where multiple QPUs are interconnected via a high-speed optical quantum network and then different schedulers for entanglement generation. Our focus is on evaluating key architectural components such as physical-layer entanglement generation, quantum data center network topology, and resource-aware scheduling under realistic constraints that influence the performance of distributed quantum execution. While other important aspects like quantum memory management, quantum error correction, and classical control signaling are crucial, they fall outside the scope of this work and are left for future exploration.

Figure \ref{fig:Clos_cross_rack_intra_rack_pair_gen} outlines the end-to-end workflow of our system. Initially, the quantum circuit is partitioned generating a DAG with local and non-local gates clearly marked. In quantum circuit compilation, the goal is to split a quantum circuit $C$, consisting of list of operations on a set of logical qubits $q = {q_0, q_1,...,q_n}$, among a set of QPUs $Q = {QPU_0, QPU_1,...,QPU_n}$ to minimize the number of inter-QPU entanglements required to execute two-qubit non-local gates. Then, a network orchestrator schedules entanglement generation for remote gate execution, considering both the circuit dependency graph and network resource constraints. In this paper, we assume a dynamic circuit-switched quantum network that efficiently distributes entanglement between QPUs by coordinating access to shared network resources, including Bell-state measurement devices, entanglement sources, and quantum memories. For intra-rack and cross-rack entanglement generation we can use different protocols as shown in the figure and we explain in detail in the next subsection.

\subsection{Intra and cross rack entanglement generation protocols}

The physical layer of the QDC is responsible for generating entangled Bell pairs (EPR pairs) between communication qubits residing on different QPUs. These EPR pairs are the fundamental resource for executing non-local quantum gates and enabling qubit teleportation \cite{eisert2000optimal}. Each QPU contains a set of data qubits (for computation) and communication qubits (for networking). Entanglement is established using photonic interfaces—either through direct photon emission (emitter-emitter schemes) or photon scattering (scatterer-scatterer schemes) \cite{monroe2014large,ferreira2024deterministic,zhan2020deterministic}. These methods produce single-photon states in encodings such as polarization, time-bin, or Fock-space \cite{ang2024arquin,monroe2014large,jiang2007distributed,bayerbach2023bell}.

For intra-rack EPR generation, the two QPUs each prepare half of an EPR pair, which are transmitted to an intermediate node where a Bell state measurement is performed to complete the entanglement. This process incurs a $50$ percent intrinsic loss due to the probabilistic nature of the BSM, in addition to fiber coupling losses that degrade the success probability exponentially with distance. The combined effect significantly reduces the overall success rate of entanglement generation, even within a single rack, and we expect realistic numbers to be around $p_{\textrm{success}}\approx0.2$. With advances in experimental techniques, we expect this number to increase. 

For cross-rack communication, the protocol follows a similar procedure but includes quantum frequency conversion \cite{van2022entangling,craddock2023highratesubghzlinewidthbichromatic}, as illustrated in Figure \ref{fig:Clos_cross_rack_intra_rack_pair_gen}. This additional step is necessary because our top-of-rack (ToR) switches and BSM devices operate in the near-infrared (NIR) spectrum, which is not optimal for long-distance transmission due to higher photon loss in standard optical fibers. To enable communication across racks, the photon emitted in the NIR band is converted into the telecom wavelength regime—which is better suited for low-loss transmission through fiber—using a bidirectional quantum frequency converter. Once the photon reaches the destination rack, it is converted back into NIR to be compatible with the local BSM hardware.

\subsection{Entanglement generation scheduling}
In this section, we investigate scheduling strategies for entanglement generation and gate execution in distributed quantum computing, where non-local gates require EPR pairs. Entanglement generation is probabilistic, and its variability plays a key role in scheduling decisions. The scheduler needs to consider gate dependencies in the circuit DAG (will be described below) and handle resource contention due to limited communication qubits and shared bell state measurement modules (BSMs). Efficient scheduling ensures EPR pairs are generated and used in a timely manner for remote gate execution or teleportation. 

To represent gate dependencies in a quantum circuit, we use a directed acyclic graph (DAG), where qubits are modeled as input and output nodes, gates are internal vertices, and edges connect inputs to outputs through the gates. In this paper, we adopt the term DAG to refer to this representation. A layer refers to a set of gates within the DAG that have no unmet dependencies and can be executed in parallel. Once all gates in a layer are executed, the next set of ready gates forms a new layer. 

Next, we describe three different scheduling strategies and discuss their advantages and limitations in the context of quantum data center architectures.

\myparab{Static Scheduling with Expected Latency:}
This approach, commonly used in prior work \cite{shapourian2025quantum}, assumes that entanglement generation completes in a deterministic time which is equal to mean entanglement generation time. This approach assumes all gates in a layer execute only after their required EPR pairs are available, based on the expected entanglement generation delay. While this approach simplifies scheduling and analysis, it overlooks randomness in entanglement generation and can underestimate execution latency under realistic, probabilistic network conditions.


\myparab{Static scheduling with probabilistic simulation}:
This approach accounts for the stochastic nature of entanglement generation, where each attempt can succeed or fail according to the given probability. The scheduler follows a strict layer-by-layer execution pattern, advancing only once all gates in the current layer have been supplied with entanglement. In this model, entanglement generation for any gate in the next layer will not begin until all gates in the current layer have completed execution, even if some of their dependent gates are already finished.

\myparab{Dynamic scheduling with probabilistic simulation}:
This approach embraces the flexibility of dynamic scheduling, where entanglement generation for a gate can begin as soon as its parent gates have completed—regardless of the gate’s original layer in the DAG. This model allows the system to opportunistically progress through the DAG, making better use of available entanglement resources and reducing idle time caused by network delays or entanglement failures. The pseudo code for this is given in Algo~\ref{alg:dynamic}.

While static scheduling with expectation values can provide a rough estimate of long-term average behavior, it is hard to capture critical real-world dynamics such as entanglement failures, queue buildups, and coherence loss due to delays in this model. Especially under constrained or variable conditions, expectation-based models can significantly overestimate system performance. Our work focuses on realistic, probabilistic models to evaluate how scheduling strategies behave under these non-ideal conditions. We developed a lightweight event-driven simulator to evaluate these scheduling strategies in a realistic QDC setup.

\begin{figure*}
\centering
    \includegraphics[width=17.6cm]{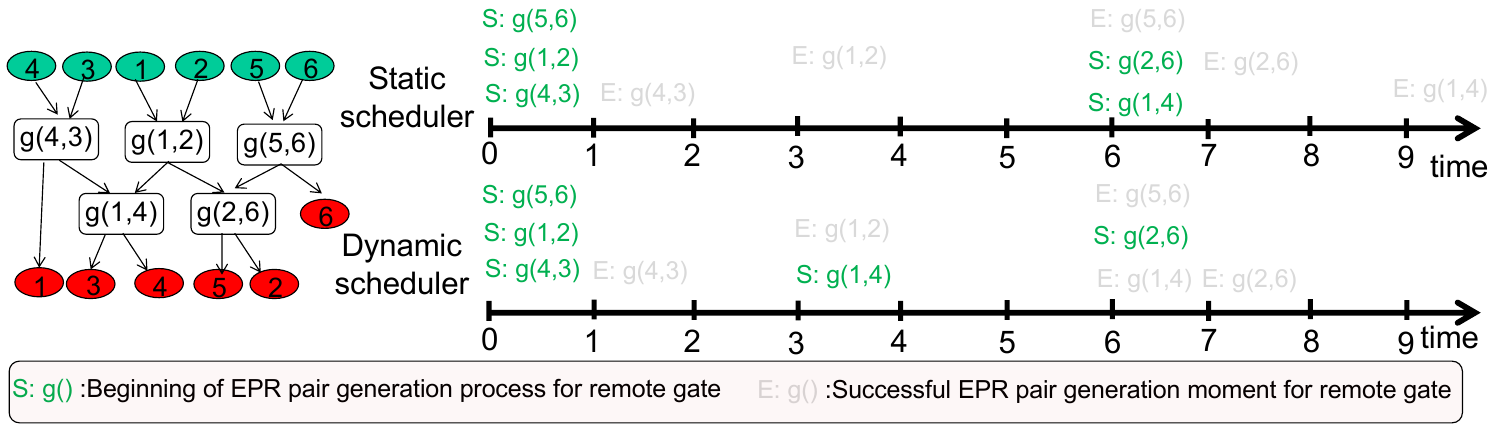}
\captionsetup{} 
  \caption{An instance of applying static and dynamic scheduling algorithm to a quantum circuit DAG and how dynamic scheduler can reduce execution delay.}\label{fig:static_dynamic_example}
\end{figure*}

Figure \ref{fig:static_dynamic_example} illustrates the behavior of static and dynamic schedulers using the corresponding DAG (shown on the left) of a circuit. We assume we only have CNOT gates between qubits in our circuit and all the CNOT gates are non-local. In a circuit, a layer is the set of gates that can be executed in parallel as they are independent gates. The subsequent panels show how remote gates are scheduled step-by-step, beginning from the frontier nodes of the DAG. The beginning of entanglement generation process for a remote gate is highlighted by green color on the time line, while the ending of this process is highlighted by gray color. Since entanglement generation is probabilistic, each gate may take different number of time units to get the EPR pair and before that multiple attempts may happen. 

\begin{algorithm}[H]
\caption{Dynamic scheduling of non-local gates}
\begin{algorithmic}[1]
\REQUIRE DAG representing the quantum circuit with annotated non-local gates
\STATE Initialize active\_tasks ← $\emptyset$
\WHILE{DAG is not empty or active\_tasks is not empty}
    \STATE front\_layer ← \texttt{DAG.get\_front\_layer()}
    \FOR{each node in front\_layer}
        \IF{node not in active\_tasks}
            \IF{\texttt{is\_two\_qubit\_gate(node)} and \texttt{resources\_available(node)}}
                \STATE active\_tasks[node] ← start\_entanglement\_generation(node)
            \ELSIF{\texttt{is\_one\_qubit\_gate(node)}}
                \STATE \texttt{DAG.remove\_node(node)}
            \ENDIF
        \ENDIF
    \ENDFOR
    \STATE finished ← wait\_for\_any(active\_tasks.values())
    \FOR{each task in finished}
        \STATE node ← \texttt{get\_corresponding\_node(task)}
        \STATE \texttt{DAG.remove\_node(node)}
        \STATE Remove node from active\_tasks
    \ENDFOR
\ENDWHILE
\end{algorithmic}
\label{alg:dynamic}
\end{algorithm}

In the given example, the initial frontier of the DAG includes CNOT gates on qubit pairs (1,2), (4,3), and (6,5), where (c, t) indicates the control and target qubits. We assume the network has sufficient resources to initiate entanglement generation for all three gates in parallel, though the completion times vary: $1$ time unit for gate $(4,3)$, $3$ time units for gate $(1,2)$, and $6$ time units for gate $(6,5)$.

Under static scheduling, the system waits until all three gates have completed entanglement generation before proceeding to the next frontier. It then begins generating entanglement for the next independent gates, $(1,4)$ and $(2,6)$, simultaneously at time step $6$. The circuit execution ends at time unit $9$.

In contrast, the dynamic scheduler updates the DAG as soon as individual gates complete. For example, once entanglement for gate $(1,2)$ is ready at time step $3$, the dynamic scheduler checks for newly available independent gates and immediately starts entanglement generation for gate $(1,4)$, even though gate $(6,5)$ has not yet completed as $(1,4)$ does not depend on $(5,6)$. Later, once entanglement for gate $(6,5)$ becomes available, the dynamic scheduler proceeds with gate $(2,6)$. This approach allows more flexibility and better utilization of  resources and the circuit execution ends at time unit $7$.





\section{Evaluation}
\label{sec:eval}
In this section, we evaluate different schedulers in our network orchestration framework where we schedule entanglement generation between different QPUs in a Clos topology to enable remote gate execution and qubit teleportation. Our goal is to show how realistic assumptions about entanglement generation, topology, and hardware constraints—such as network reconfiguration delay, probabilistic entanglement generation, and communication qubit coherence time affect circuit execution delay in a QDC. The primary metric is the average total time required to generate EPR pairs for remote gates after circuit partitioning, with total latency measured from start to completion of all gates considering the dependency between gates and network constraints. We assume local gate times are negligible and we only care about the time it takes to generate entanglements.

We use the WBCP algorithm proposed in \cite{kaur2025optimized} as our circuit partitioning algorithm and evaluate different schedulers using a diverse set of circuits, including Quantum Fourier Transform (QFT), Quantum Volume (QV), QAOA, and Random circuits—each with 100 qubits to represent large-scale quantum workloads. These circuits are decomposed into a native gate set (cx, h, rz, crz) using Qiskit's \texttt{transpile()} function to ensure compatibility with realistic hardware constraints. We incorporate realistic hardware assumptions: emitter-emitter protocol with Fock space encoding for intra-rack communication, and scatterer-scatterer protocol for cross-rack entanglement generation. 

\begin{figure}
\centering
    \includegraphics[width=6.4cm]{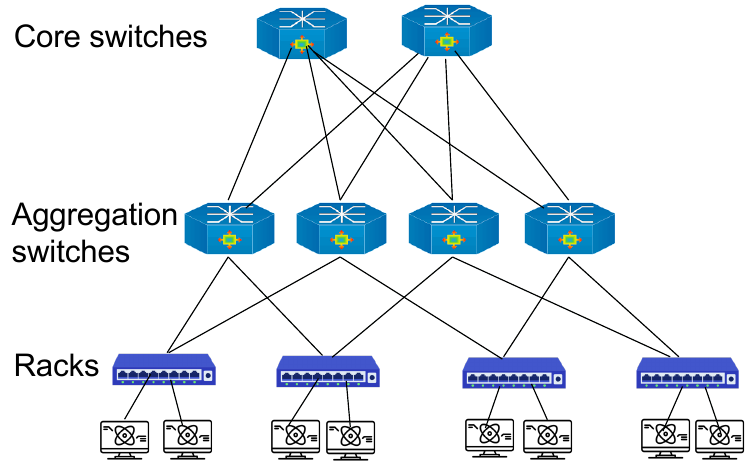}
\captionsetup{justification=centering} 
\vspace{-0.05in}
  \caption{Quantum data center topology used in the evaluation.}\label{fig:clos_evalaution}
\end{figure}

We adopt a Clos topology with 2 core switches both connected to 4 aggregation switches and we have 4 racks each with 2 QPUs as our QDC network. Each aggregation switch is connected to two racks. With this setup, we have 8 QPUs. The topology is shown in figure \ref{fig:clos_evalaution}. Each rack Intra-rack entanglement generation success probability is $0.5$ and cross-rack EPR pair generation success probability is variable, defaulting to $0.2$ unless otherwise stated. In additions, we assume the intra-rack entanglement attempt time is $1$ microseconds, while cross-rack attempt time is $10$ ms. Optical switch reconfiguration delay is assumed to be $1$ ms.

We explore two gate execution strategies (static and dynamic scheduling with probabilistic simulation) explained in section \ref{sec:schedulers}. For each gate, the system checks whether an entangled path is available using shortest-path routing. If sufficient resources are available—meaning each involved QPU has at least one free communication qubit and at least one BSM is available on the switches along the path—then those resources are reserved for that QPU pair until the EPR pair is successfully generated for remote gate execution. If the required resources are not available, the gate’s execution is deferred until a valid entangled path becomes available. We run the simulation 100 times for each setup and get the average circuit execution delay.

\subsection{Impact of scheduling strategies on circuit execution delay}
\begin{figure*}
\centering
    \includegraphics[width=18.7cm]{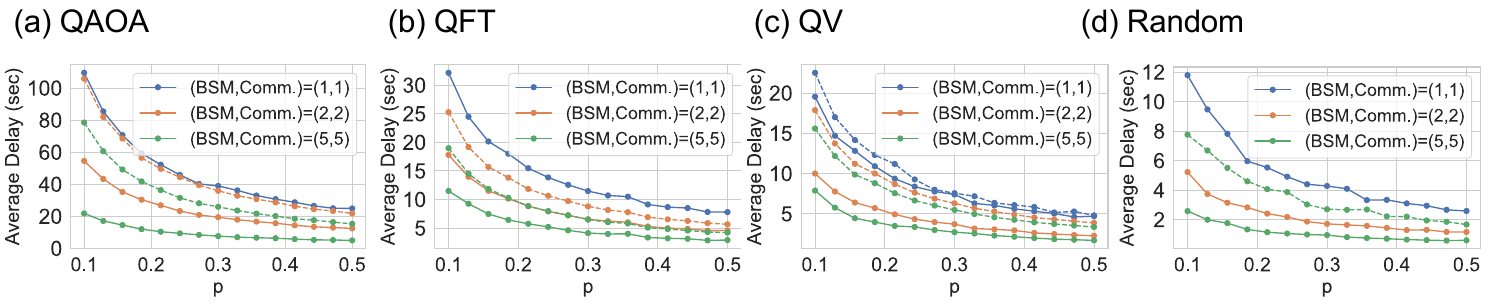}
\captionsetup{justification=centering} 
\vspace{-0.23in}
  \caption{Execution delay vs. cross-rack entanglement success probability ($p$). Solid lines: dynamic scheduler. Dashed lines: static scheduler. Parentheses show (BSMs, communication qubits) in a 4-QPU setup. Intra-rack success probability is fixed and is $\frac{1}{2}$.}\label{fig:delay_as_p}
\end{figure*}

\begin{figure}
\centering
    \includegraphics[width=9cm]
    {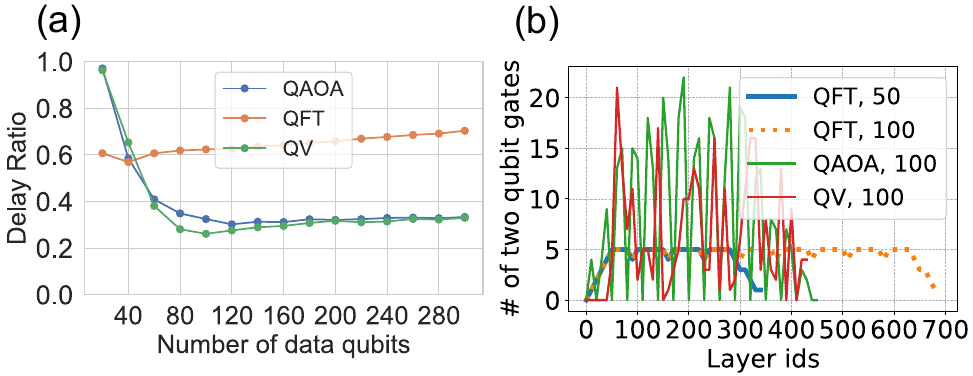}
    \vspace{-0.23in}
  \caption{The ratio of dynamic scheduler delay to static scheduler delay in QAOA, QFT, and QV circuits with different number of data qubits (a) and number of two qubit gates in each layer of different circuits (b) Our setup has 4 QPUs and 5 BSMs at each switch. }\label{fig:ratio_as_bw_bsm_qubits}
\end{figure}

Figure \ref{fig:delay_as_p} shows the average circuit execution delay as a function of cross-rack entanglement generation success probability ($p$), comparing static (dashed lines) and dynamic (solid lines) schedulers across different numbers of communication qubits and BSMs. As expected, the dynamic scheduler reduces delay by improving resource utilization and not delaying gates whose parents are already executed. However, not all circuits benefit equally. For example, the dynamic scheduler shows less improvement for QFT compared to QAOA and QV. This is because each layer in the QFT circuit's DAG contains only a few non-local gates, limiting opportunities for parallel entanglement generation. In contrast, circuits like QAOA, which have more two-qubit gates per layer (Figure \ref{fig:ratio_as_bw_bsm_qubits}b), offer more opportunities for the dynamic scheduler to outperform the static approach.

We next analyze the performance gain of the dynamic scheduler over the static scheduler under varying system parameters. Figure \ref{fig:ratio_as_bw_bsm_qubits} shows the ratio of average execution delay (dynamic vs. static) as a function of circuit size (number of data qubits). As expected, the ratio decreases with larger circuits, indicating that the dynamic scheduler benefits from greater parallelism in EPR pair generation. However, for circuits like QFT, the ratio remains close to 0.65 even for large sizes, due to their limited number of non-local gates per layer. In contrast, circuits like QAOA show significant improvement with dynamic scheduling.

\subsection{Lookahead strategy with memory cut-off}

We now evaluate the impact of communication qubit cutoff time on circuit execution delay. In this simulation, we assume a lookahead strategy for the network orchestrator. After initiating entanglement generation for the current frontier gates in the DAG, the orchestrator proactively attempts to generate entanglement for gates in upcoming DAG layers, provided that network resources are available. If successful, the resulting EPR pairs are stored in memory by the communication qubits until they are needed. This is in contrast to previous scheduling strategies (without lookahead), where EPR pairs were immediately consumed upon generation. With the lookahead strategy, EPR pairs may be held for a short duration, which introduces the risk of decoherence. To model this, we define a cutoff time—the maximum duration a communication qubit can store an EPR pair. If the EPR pair is not used before this time expires, it is discarded. In our simulation, the cutoff time begins from the moment the entanglement is successfully generated, marking the start of the EPR pair's lifetime in the communication qubit.

\begin{figure}
\centering
    \includegraphics[width=6cm]
    {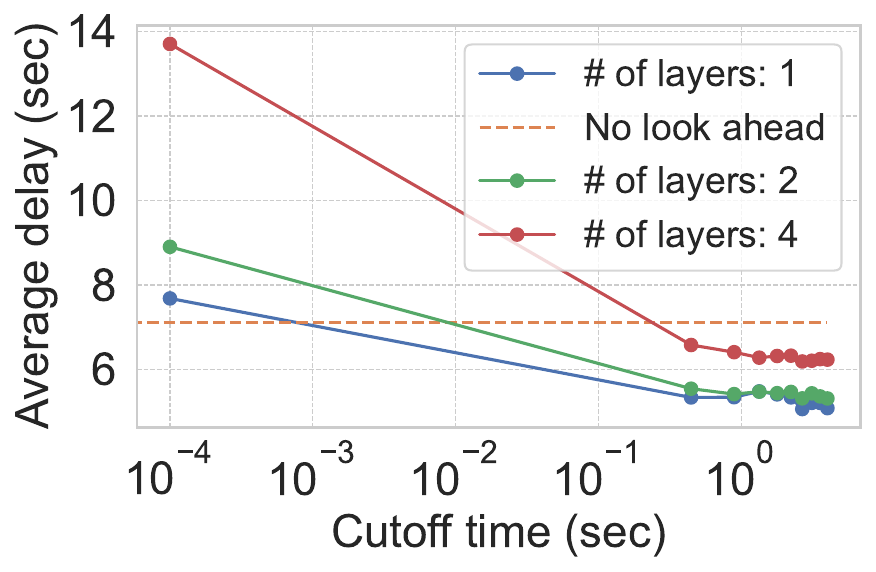}
\captionsetup{justification=centering} 
  \caption{Execution delay vs. communication qubit cutoff time. ``$\#$ of layers'' indicates how many future DAG layers the orchestrator targets for lookahead entanglement generation.}\label{fig:dealy_as_t_coh}
\end{figure}

Figure \ref{fig:dealy_as_t_coh} shows the average execution delay for a QAOA circuit with 100 qubits distributed across 4 QPUs, as a function of cutoff time. As shown, short cutoff times do not reduce mean execution delay and may even degrade performance. This is because entanglement generated in advance for gates in upcoming layers is discarded before it can be used, leading to wasted resources and preventing the network from serving gates in the current DAG layer. However, with a sufficiently high cutoff time, the lookahead strategy becomes beneficial, enabling the orchestrator to reduce total execution delay. As a potential direction for future work, one could explore a protocol that immediately initiates new entanglement generation when previously stored EPR pairs are discarded due to coherence limits. Such a reactive approach could help maintain high resource utilization and further improve overall performance.

\subsection{Average execution delay vs resources:}
\begin{figure*}
\centering
    \includegraphics[width=17.5cm]
    {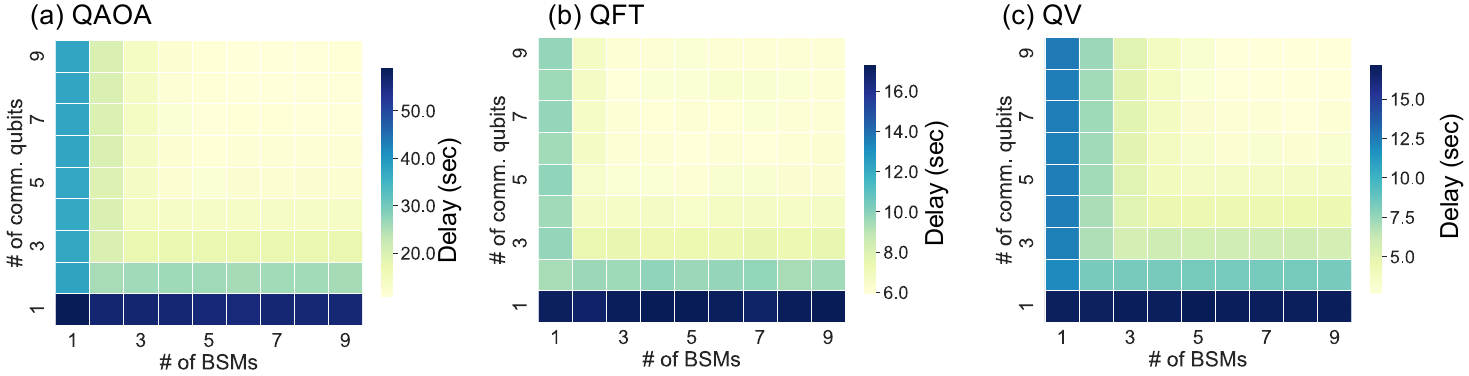}
\captionsetup{justification=centering} 
  \caption{Circuit execution delay as a function of resources for 4 QPUs and 100 qubits in each circuit.}\label{fig:dealy_as_bw}
\end{figure*}

We now focus on the effect that network resources have on average circuit execution delay. Figure \ref{fig:dealy_as_bw} illustrates how average delay varies with the number of communication qubits per QPU and the number of BSM modules per switch for different circuits. We have used 4 QPUs and 100 data qubits for each circuit. The scheduling strategy is the dynamic scheduling without lookahead strategy. While increasing either resource helps reduce delay, the benefit plateaus beyond a certain point. Specifically, if QPUs have too few communication qubits, additional BSMs cannot be fully utilized—and conversely, more communication qubits offer limited benefit if BSMs become the bottleneck.

\subsection{Congestion-free network}
In this experiment, we assume an idealized setting where each switch in the quantum data center is equipped with an unlimited number of BSM modules and we have an unlimited number of communication qubits available. This setup allows us to observe the true BSM demand patterns over time without any contention or resource limitations. Our goal is to measure, for each switch and each circuit, how many BSMs are required every moment during circuit execution to support congestion-free operation.

For benchmark circuits QAOA and QV, we simulate distributed execution over a Clos-topology network with a fixed number of QPUs (8) and unlimited communication qubits and BSMs. At every time step, we record the maximum number of BSMs actively in use across all switches, reflecting the level of parallel entanglement generation required by the orchestrator. This allows us to construct a BSM demand profile for each circuit, capturing both the peak usage across switches and the temporal evolution of entanglement load across the network. 

\begin{figure}
\centering
    \includegraphics[width=9cm]{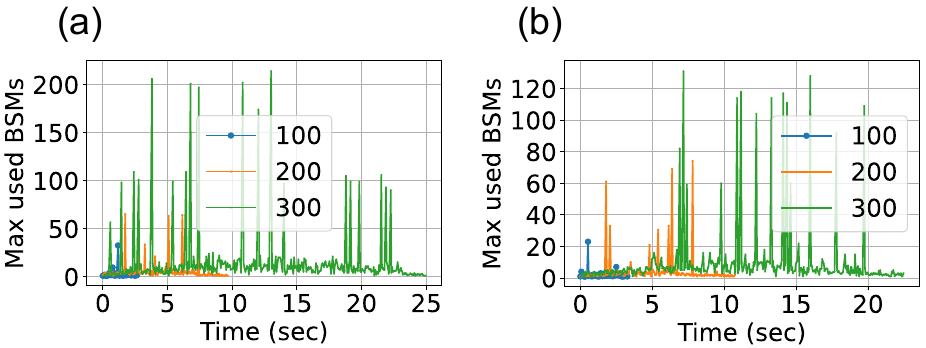}
  \caption{Maximum number of used BSMs at each time step for one run of a QAOA circuit with different number of data qubits on 4 QPUs (a) and 8 QPUs (b) with unlimited resources.}\label{fig:max_BSM_used_over_time}
\end{figure}

Figure \ref{fig:max_BSM_used_over_time} shows the maximum number of BSMs used across all switches over time during the execution of QAOA circuits with varying numbers of data qubits, for both 4-QPU (a) and 8-QPU (b) configurations. Rather than reporting BSM usage at fixed time intervals (e.g., every millisecond or microsecond)—which would require choosing an arbitrary time granularity and introduce additional monitoring overhead—we instead sample usage each time the orchestrator reaches line 19 of Algorithm~\ref{alg:dynamic}, i.e., when it completes an attempt to serve the current DAG front layer., we sample usage each time the orchestrator reaches line 19 of Algorithm~\ref{alg:dynamic}, corresponding to when it completes an attempt to serve the current DAG front layer. The results show that supporting congestion-free execution of a 300-qubit QAOA circuit on 4 QPUs requires approximately 200 BSMs per switch. Expanding to 8 QPUs reduces this requirement to about 120 BSMs, as the wider network distributes entanglement load across more switches, reducing contention. 

\begin{figure*}
\centering
    \includegraphics[width=18.3cm]{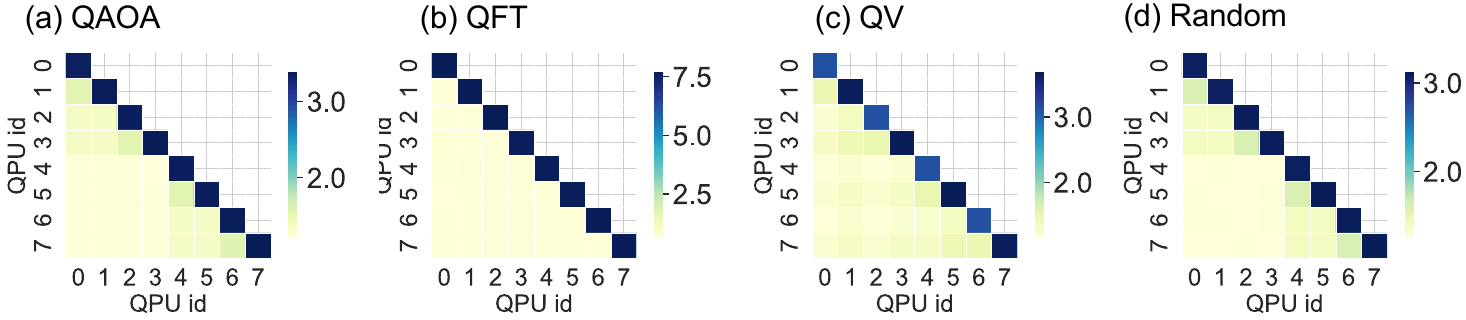}
    \vspace{-0.06in}
\captionsetup{justification=centering} 
  \caption{Heatmap of QPU-to-QPU entanglement generation for different circuits. Each circuit produces a distinct communication pattern. The color bar indicates the percentage of total entanglement generated between all pair of QPUs.}\label{fig:heat_map_demand}
\end{figure*}

Finally, Figure~\ref{fig:heat_map_demand} presents a heatmap showing the distribution of entanglement requests (both intra and cross-rack) between QPU pairs for each circuit. This demand pattern is circuit-dependent and can inform the design of future entanglement schedulers, similar to traffic-aware scheduling in classical data centers \cite{zerwas2024d3}. Most requests are concentrated between a subset of QPU pairs and also along the diagonal, indicating frequent communication within the same QPUs. This is expected, as the circuit partitioning algorithm tends to assign qubits with frequent two-qubit interactions to the same QPU—resulting in more local gates. A more advanced, network-aware partitioning algorithm could further optimize this by assigning qubits not only to the same QPU but also to QPUs within the same rack, thereby increasing the proportion of intra-rack entanglement, which is more efficient to generate than cross-rack entanglement.


\section{Conclusion}
\label{sec:conclusion}
In this work, using an event-driven simulation framework, we evaluated static and dynamic scheduling strategies over a reconfigurable photonic-switch-based network with probabilistic entanglement generation, limited communication qubits, and constrained Bell-state measurement (BSM) resources for a QDC. Our results demonstrate that dynamic scheduling offers significant advantages over static scheduling in reducing circuit execution delay, particularly for circuits with high gate parallelism, such as QAOA. We also introduced a coherence-aware model using communication qubit cutoff times and showed that aggressive lookahead scheduling can degrade performance when coherence is limited. Our findings provide practical insights into how network topology, resource distribution, and gate scheduling strategies jointly impact the performance of distributed quantum computing in a QDC. This work takes a step toward bridging circuit-level compilation with system-level orchestration, enabling more efficient and scalable deployment of quantum workloads across modular quantum architectures.

\bibliography{refs.bib}
\end{document}